# Influencing Students' Relationships With Physics Through Culturally Relevant Tools

Ben Van Dusen and Valerie Otero

*School of Education, University of Colorado, Boulder, 80309, USA*

**Abstract.** This study investigates how an urban, high school physics class responded to the inclusion of a classroom set of iPads and associated applications, such as screencasting. The participatory roles of students and the expressions of their relationships to physics were examined. Findings suggest that iPad technology altered classroom norms and student relationships to include increased student agency and use of evidence. Findings also suggest that the iPad provided a connection between physics, social status, and play. Videos, observations, interviews, and survey responses were analyzed to provide insight into the nature of these changes.

## INTRODUCTION

On January 12$^{th}$, 2012, two high school physics students took the initiative to leave the classroom with an iPad to shoot a video in their car. The video showed them starting to drive, placing a coffee cup on their dashboard, and then quickly accelerating to a stop. The video clearly shows that the coffee cup kept moving forward even after the car had come to rest. The students inserted this video into their digital lab report and showed it to their classmates.

The above example struck us as a particularly intriguing interaction. How can the actions of the students be explained? What compelled the students to stretch the bounds of the classroom in ways that were never formally sanctioned by the teacher in order collect and log data from a relevant experiment that they designed?

The iPad Enhanced Active Learning (iPEAL) project was designed specifically to explore the effects iPads in high school physics classrooms. Like the personal computers (PCs) before them, iPads have been hyped as a "magical" product that will revolutionize education. PCs introduced new ways of collecting and analyzing classroom data through probeware [1], video-based motion analysis [2], and introducing model-like evidence through simulations [3]. Such PC-based activities have changed the role of data collection and analysis in classroom physics. At the same time, specific populations of students remain largely underrepresented in university and college physics. We hypothesize that in order to capture the natural curiosity of *all* students, and to introduce them to the richness that physics inquiry could bring to their lives, the classroom environment must be shifted significantly to facilitate the gradual process of personal identification with physics.

By examining changes in student activities and peer-interactions, we investigate the questions: (1) In what ways do iPads change student interactions with physics, if at all? (2) In what ways do iPads mediate student relationships to physics, if at all?

## RESEARCH CONTEXT

This research was conducted in 5 high school physics classes (4 regular and 1 Advanced Placement) in an urban area. The school is primarily composed of students who have been traditionally underserved and are underrepresented in science. While the student enrollment of the courses varied throughout the year, there were approximately 140 students at any given time. At the start of the school year 73% of the students were juniors and 27% were seniors.

The five classes shared a single set of 38 iPads. This allowed each student to have an iPad that was unique to them during class, but was shared with four other students throughout the day.

A third-year teacher with a background in biology, including a Ph.D. in biochemistry, taught all of the classes. Like most high school teachers of physics, she did not have a physics or physics education degree [4]. She is a Streamline to Mastery [5] teacher, engaged in NSF-funded teacher-driven professional development.

Students engaged in iPad-supported activities that were intended to supplement traditional physics assignments. For example, students created screencasts of their textbook problem solutions. Through screencast technology, they created a video of the iPad screen as they recorded think-aloud audio while solving physics problems using the stylus. This allowed students to record and play back their dynamic problem solutions. These screencasts were later made available to other students.

## THEORETICAL FRAMEWORK

We take a critical perspective, where we assume that high school student physics experiences are too often wrought with fear and failure rather than being enjoyable, empowering, and personally meaningful. Our research is based on the assumption that in order for learning to occur, the learner must be engaged in an activity that is personally meaningful. Further, for an activity to be personally meaningful, it must produce, or be produced by, positive experiences. We focus on the iPad as a tool that could potentially *mediate* a change from negative to positive experiences in physics. From these assumptions we have created a tentative model (Fig. 1) in which positive experiences and personally meaningful activities are both reflexive and necessary for creating an environment in which learning may occur.

While similar models have been described in the literature [6], we were drawn to this type of model through our first year of observations in the iPad learning environment. The construct of "personal meaning" was salient in the social context yielding potential for helping us understand the experiences that students were having with physics via the iPad. During the second-year of observation and interviews we intend to further articulate this model in terms of the specific ways in which students engage with the iPad, physics, or both.

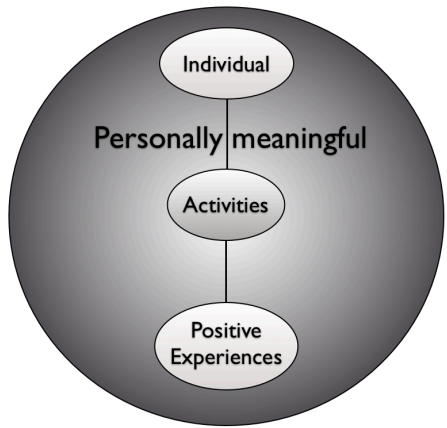

**FIGURE 1.** A simple model of personally meaning.

## METHOD

Field-notes, video recordings, artifacts, student surveys, and student interviews were collected. The interviews were typically administered the same week as the surveys and included a similar set of question in order to collect expanded answers from a smaller set of students. Findings from the interviews and surveys were triangulated with the field-notes from the weekly observations. Classroom videos were used as references to supplement field-notes. We used a generative coding methodology to discover any themes that emerged from the data. Eleven themes were found, four of which are relevant to the construct of personal meaning. The use of iPads in this context: 1) facilitated student use of evidence, 2) facilitated student play, 3) increased student agency, and 4) appeared to impact student social status. These findings along with exemplars are described in the section below.

## FINDINGS

**Finding 1:** The iPads facilitated students' use of evidence. This was accomplished by making it easy to complete tasks that were previously either difficult or impossible. By facilitating data collection, analysis, and collaboration, the iPads allowed students to draw their own conclusions based on evidence, rather than relying on the book or the teacher to provide solutions.

The shift of authority from the teacher and textbook to evidence was most apparent during labs. For example, the iPad altered the task of evaluating the results of a sound lab. The original lab required students to cut PVC piping to predetermined lengths in order to produce different harmonic frequencies. Students then listened to the pitch of the pipe and were to determine if they had created the intended pitch. Only the teacher and one of the students in the observed classes had the musical training to identify the pitch classes by ear. The rest of the students were unable to aurally identify if their pipes were producing the correct frequency and relied on the teacher's assessment. With the introduction of the iPad, students used several applications to identify the pitches of the PVC experiment. One application displayed the frequency the pipes were emitting numerically, another produced varying pitches for students to reference and match, and the last application identified the pitch class and reflected the tuning variations on a virtual dial. With these three applications, the iPad *transformed the task* of asking the teacher to evaluate the sound into a different task in which students could use a mathematical, audio-matching, or visual-spatial model to assess the sound. There were many situations such as this, where tasks were transformed from teacher-as-authority to evidence-as-authority. These transformations were tractable for students and allowed them to reason with evidence instead of deferring to an external authority.

**Finding 2:** The iPad facilitated student play. Even before the iPads were actually implemented, students demonstrated excitement to use them in class. During the first two months of the school year, when students did not yet have the iPads, they regularly inquired about when the iPads would be arriving. Once the iPads were implemented, students used the time before class and during transitions to explore the iPads. Typical student-initiated activities included taking pictures of friends, setting the background image, exploring simulations, and playing games. Unlike previous years, student began to regularly go into the physics classroom outside of class time to work on physics projects. When asked how much they enjoyed doing iPad work versus traditional work, students articulated a strong preference for iPad work (Fig. 2). The students who found the work to be less enjoyable on the iPad, expressed a preference for laptops.

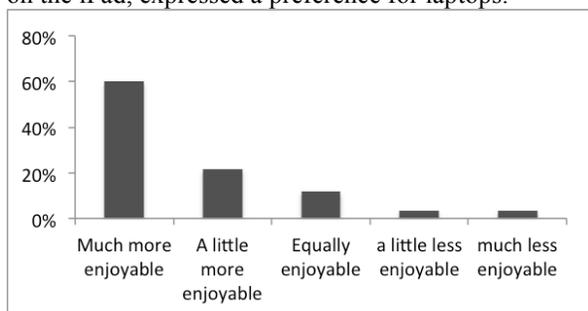

**FIGURE 2.** Comparison of iPad to traditional work.

**Finding 3:** The iPad increased student agency. Through specific activities made possible by the iPad, such as students' creation of screencasts, some responsibilities were shifted from the teacher to the student. As described earlier, students used screencast technology to record a verbal explanation with video documentation of the steps in their problem solutions and then shared them with other students. When the AP students were asked whether the teacher, themselves, or the class determined what steps should be shown in their work, students were more likely to reference the teacher for traditional lab work and themselves for screencasts (Fig. 3), none said the class.

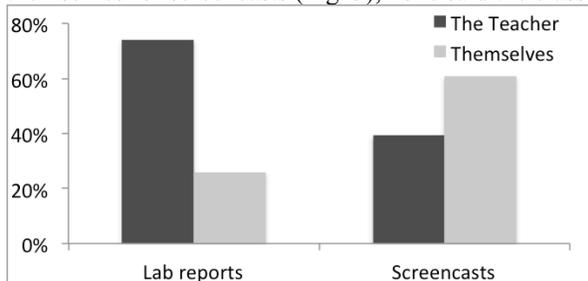

**FIGURE 3.** Who determines the steps to be shown.

Students were aware of and able to vocalize the differences in personal agency in determining steps in physics labs. When asked if screencasts helped them learn, one AP student answered, "You can learn from visuals and reading, but teaching, writing, and teaching yourself again is a very effective way. I think it's the most effective way because you think that you're going to give someone a lesson and you test your own knowledge. You don't have anyone telling you, you test yourself" (Manuel, 5/4/12). Another AP student answered, "Trying to explain to someone else is like being a teacher" (Julia, 5/4/12). We infer from this set of data that the iPad has the potential to increase student agency more broadly through activities such as self-guided screencasts.

**Finding 4:** The iPad appeared to impact student social status. We have some evidence that has led us to further investigate this claim. It was common for students in non-physics classes to express a desire to be involved with the iPad classroom environment. When physics students were asked what they enjoyed about using the iPad, one student said, "I do really like the photobooth, it's really cool. I can send pictures to my email and put them up [on Facebook]. Then that's another way for people to know that we have iPads in our class. They're like, 'how did you do that?' I'm like, 'oh we have iPads in our physics class,' and they're like, "what?!" (Sally, 1/13/12)

Using the iPads was important to the students. For example, when they learned that there was not enough money to purchase AppleCare, cases, and styluses for the iPads, the students offered to raise the money or pay for the equipment themselves.

The effect of the iPads reached beyond the class to raise the status of the school itself. The physics classes had more iPads than the rest of the district combined. Once the iPads were implemented, school visitors were brought to observe the physics classes on a regular basis. One prospective student was overheard by the teacher saying that he thought the school was "ghetto" and was not going to go there until he heard that it had "the physics classes with iPads."

Data collection to support the claim of social status has been challenging, but we intend to continue to investigate this and our other claims throughout the upcoming year.

For research purposes, we have analytically separated play, status, and agency but some findings were difficult to categorize. For example, when asked what percentage of their digital assignments (e.g. screencasts and digital lab reports) and analogue assignments (e.g. book problems and handwritten labs) they shared with fellow physics students, AP and regular students jointly expressed that they shared more of their digital assignments (Fig. 4). We were uncertain how this finding should be classified; it could represent any number of things ranging from agency and status to teacher-driven requirements. We report it here because it reveals a possible shift in the classroom collaborative environment.

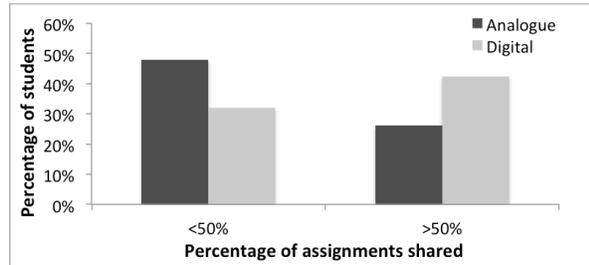

**FIGURE 4.** Percentage of assignments shared with fellow physics students.

## THEORETICAL IMPLICATIONS

Based on our initial findings, we have further operationalized the construct of positive experiences to include social status, play, and agency (Fig. 5). Using this lens we can investigate the potential role of the iPad in physics learning. We postulate that the iPad created a "bridge" connecting students, via social status, play, and agency to physics.

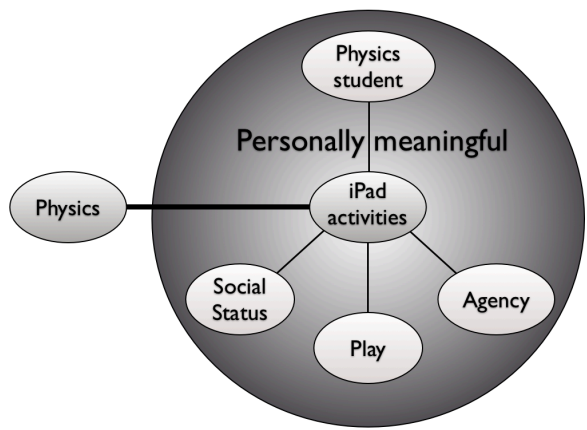

**FIGURE 5.** Personal meaning connects students to physics.

We propose a model in which the iPad may have the potential to mediate a positive relationship between students and physics. Based on our current data analysis, we suggest that agency, status, and play afforded by the iPad can serve as mechanisms for reshaping students' relationships with physics.

Although this argument hedges on classical conditioning/behaviorist models of learning appearing in the literature as early as 1910 [7], it may have some value in helping to connect what a student internally experiences in the physics classroom to the external experiences that are provided in the class environment. These experiences may produce fear and anxiety or they may feel empowering and exciting to the student. We provided some evidence that a mediating artifact such as the iPad could serve to facilitate a positive relationship with physics through personally meaningful activities that engender a sense of play, status, and agency. Our model goes further in positing that the "bridge" between physics and the iPad could eventually be switched off and student's sense of personal engagement in physics would remain (Fig. 6). In such cases, the iPad has *mediated* the student's positive relationship with physics. It follows that the student would be more likely to continue to engage in future studies of physics.

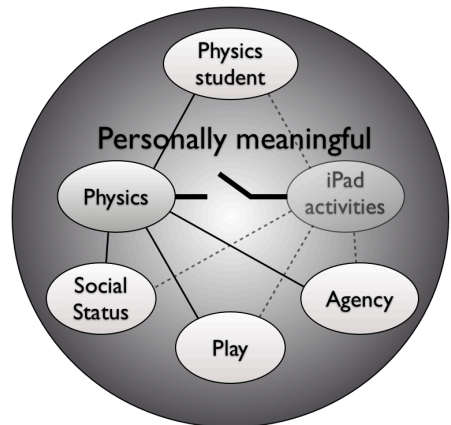

**FIGURE 6.** Shifted student relationship with physics.

It is unknown if any of the students in the study reported here fell into this category. However, the opening example of students videotaping and testing Newton's first law provides some indication that such a transformation could be taking place. We will continue to test this model in the future, collecting data that is specifically relevant to the categories that were revealed through the current study.

## ACKNOWLEDGEMENTS

We thank Susie Dykstra for creating, and allowing us to study, her iPad physics enhanced environment, and the WISE and NSF (DUE #934921) grants.

## REFERENCES

1. R. Thornton & D. Sokoloff, *Am. J. Phys* **66**, 338 (1998)
2. P. Laws, *Phys. Today*, December, 24 (1991)
3. D. Clark, B. Nelson, P. Sengupta, & C. D'Angelo. *Learning science: Computer games, simulations, and education workshop sponsored by the National Academy of Sciences,* 36 (2009)
4. T. Hodapp, J. Hehn, and W. Hein. *Phys. Today*. 40 (2009)
5. B. Van Dusen & V. Otero. *PERC proceedings 2011*, 375 (2011)
6. H. Spencer, *Education: Intellectual, Moral, and Physical* (1860)
7. E. Thorndike, *Journal of Ed. Psych*, **1**, 5 (1910).